\documentstyle[aps,twocolumn]{revtex}
\begin{document}

\newcommand{\be}{\begin{equation}}
\newcommand{\ee}{\end{equation}}
\newcommand{\bea}{\begin{eqnarray}}
\newcommand{\eea}{\end{eqnarray}}
\newcommand{\nn}{\nonumber}

\title{Hamiltonian Reduction of $SU(2)$ Yang-Mills Field Theory}
\author{A. M. Khvedelidze \ $^a$
\thanks{Permanent address: Tbilisi Mathematical Institute,
380093, Tbilisi, Georgia.} and
\,\, H.-P. Pavel\ $^b$ }
\address{$^a$ Bogoliubov Theoretical Laboratory, Joint Institute for Nuclear
Research, Dubna, Russia}
\address{$^b$ Fachbereich Physik der Universit\"at Rostock,
              D-18051 Rostock, Germany}
\date{\today}
\maketitle

\begin{abstract}

The unconstrained system equivalent to $SU(2)$ Yang-Mills 
field theory is obtained
in the framework of the generalized Hamiltonian formalism using the method
of Hamiltonian reduction. The reduced system is expressed in terms of 
fields which transform as spin zero and spin two under spatial rotations.

\end{abstract}

\bigskip

\bigskip

\pacs{PACS numbers: 11.10.Ef, 11.15.-q, 11.15.Me}

\bigskip
\bigskip

The  degenerate character of the conventional Yang-Mills action
for \(SU(2)\)  gauge fields \( A^a_{\mu}(x) \)
\be \label{eq:act}
S[A] = - \frac{1}{4}\ \int d^4x\ F^a_{\mu\nu} F^{a\mu \nu}
\ee
leads to a restriction of the corresponding phase space spanned by 
the canonical variables 
$(A^a_0,  P^a_0) $ as well as $(A_{ai}, E_{ai})$
due to the  primary constraints $P^a_0 (x) = 0~$
and the secondary constraints, the non-Abelian Gauss law  
\bea  \label{eq:secconstr}
&&\Phi_a : = \partial_i E_{ai}+g\epsilon_{abc} A_{ci} E_{bi} = 0\ ,
\eea  
Since they are first class 
\be
\label{eq:secconstr2}
\{\Phi_a (x) , \Phi_b (y)\} =  g\epsilon_{abc}\Phi_c \delta(x-y)~,
\ee
the dynamics of the system is not uniquely predictible. 
The main problem in the Hamiltonian formulation of Yang-Mills theories
is to find the projection from the 
initial phase to the  phase space of 
unconstrained  gauge invariant variables with uniquely predictable 
dynamics. 
The conventional perturbative gauge fixing method \cite{FadSlav}
for solving this problem works successfully for the description of 
high energy phenomena,  but fails in applications in the infrared region. 
The correct nonperturbative reduction of gauge theories 
\cite{GoldJack}-\cite{KhvedZero}, on the other hand, leads to 
representations for the unconstrained Yang-Mills systems
which are valid also in the low energy region
but unfortunately are very complicated for practical calculations.
The problem is to state some practical form of the 
theory preserving all main properties of initial gauge theory 
which can applied directly to the solution of infrared problems.
With this aim we follow the method of Hamiltonian reduction (\cite{GKP} 
and references therein)
in the framework of the the  Dirac constraint formalism \cite{DiracL,HenTeit}.
In previous work \cite{GKMP} devoted to the case of the 
mechanics of spatially constant $SU(2)$ Dirac Yang-Mills 
fields we obtained the corresponding
unconstrained system desribing the dynamics of a symmetric second 
rank tensor under spatial rotations.

In this letter we  generalize our approach to field theory. 
We give a Hamiltonian formulation of classical $SU(2)$ Yang-Mills 
field theory entirely in terms of gauge invariant variables. 

The non-Abelian character of the secondary constraints (\ref{eq:secconstr2})
is the main obstacle for the corresponding projection to the
unconstrained phase space.
The way to avoid this difficulty is to replace the non-Abelian
constraints (\ref{eq:secconstr2}) by a new set of Abelian constraints
$\Psi_\alpha$ which describe the same constraint surface\footnote{
There are known several methods of the Abelianization of constraints
(see  e.g. \cite{GKP,HenTeit}   and   references therein).}.
For the new Abelian constraints $\Psi_\alpha$
the projection to the reduced phase space
can be simply achieved in the following two steps. One performs a canonical 
transformation to new variables such that part of the new momenta 
${\overline P}_\alpha$
coincide with the constraints $\Psi_\alpha $. After the projection onto the
constraint shell, i.e. putting in all expressions
${\overline P}_\alpha = 0$, the coordinates canonically conjugate to the 
${\overline P}_\alpha$ 
drop out from the physical quantities. The remaining 
canonical pairs are then gauge invariant and form the basis for the 
unconstrained system.

The problem of Abelianization is considerably simplified when
studied in terms of coordinates adapted to the action of the gauge group.
The knowledge of the local gauge transformations
of the Yang-Mills action (\ref{eq:act}), 
$A_\mu \,\,\, \rightarrow \,\,\,  A_\mu^{\prime}   =
U^{-1}(x) \left( A_\mu  - {1\over g}\partial_\mu \right) U(x)$,
directly promts us with the choice of adapted coordinates by using the
following point transformation to the new set of Lagrangian coordinates
$ \overline{Q}$  and ${Q^\ast} $
\be 
\label{eq:gpottr}
A_{ai} \left(\overline{Q}, Q^{\ast} \right)=
O_{ak}\left(\bar{Q}\right) Q^{\ast}_{ki}
- {1\over 2g}\epsilon_{abc} \left(O\left(\overline{Q}\right)
\partial_i O^T\left(\overline{Q}\right)\right)_{bc}~,
\ee
where \( O  \) is an orthogonal matrix  and \( Q^\ast \) is a
positive definite symmetric matrix.
\footnote{The freedom to use other canonical variables 
in the unconstrained phase space corresponds to another fixation 
of the six variables $Q^\ast$ in the representation (\ref{eq:gpottr}). 
This observation clarifies the connection 
with the conventional gauge fixing method. We shall discuss this point
in forthcoming publications (see also ref. \cite{Muz}).}
The transformation (\ref{eq:gpottr}) induces a point canonical transformation
linear in the new canonical momenta \( P^{\ast}_{\ ik} \) and
\(\overline{P}_{i} \). Using the corresponding  generating functional
depending on the old momenta and the new coordinates
\be
F_3 \left[ E; \ \overline{Q}, Q^{\ast}\right] :=  \int d^3z \ E_{ai}(z)
A_{ai} \left(\bar{Q}(z), Q^{\ast}(z)\right)~
\ee
one can obtain the  transformation to new canonical momenta
\( \overline{P}_{i} \) and \( P^{\ast}_{\ ik} \)
\bea  \label{eq:mom1}
\overline{P}_j (x)& :=  & \frac{\delta F_3 }{\delta \overline{Q}_j(x)}
= - \frac{1}{g} \Omega_{jr}\left(D_i(Q^{\ast})O^TE\right)_{ri}     ~,\\
\label{eq:mom2}
P^{\ast}_{\ ik}(x)& :=  &\frac{\delta F_3}{\delta Q^{\ast}_{ik}(x)}
= \frac{1}{2}\left(E^TO + O^T E \right)_{ik}~.
\eea
Where \(
\Omega_{ji} \, : = (1/2)\epsilon_{lim}\left(
O^T\left(\overline{Q}\right)\partial O \left(\overline{Q}\right)/
\partial \overline{Q}_j\right)_{lm}
\)
is assumed to be invertible matrix and 
$D_i(Q^{\ast})$ is the corresponding covariant 
derivative in the adjoint representation 
\(
\left(D_i(Q^{\ast})\right)_{mn} := \delta_{mn}\ \partial_i +
g \epsilon_{mkn}\ Q^{\ast}_{ki}.
\)
A straightforward calculation based on the  linear
relations  (\ref{eq:mom1}) and (\ref{eq:mom2})
between old and new momenta
leads to the  the following expression for the field strengths $E_{ai}$
in terms of the new canonical variables
\be  \label{eq:elpotn}
E_{ai} = O_{ak}\left( \overline{Q}\right) \biggl
[\,  P^{\ast}_{\ ki} +\epsilon _{kis}
 {}^\ast D^{-1}_{sl}(Q^{\ast})
\left[
\left(\Omega^{-1} \overline{P} \right)_{l} -
{\cal S}_l\,\right]\,\biggr]~.
\ee
Here  \( {}^\ast D^{-1}\) denotes  the inverse of the matrix operator
\(
{}^\ast D_{ik}(Q^{\ast}) : = {1\over 2}\epsilon_{imj} D_m(Q^\ast)_{jk}
\)
and
\be
\label{eq:spin} 
{\cal S}_k (x) := \epsilon_{klm}\left(P^\ast Q^{\ast}\right)_{lm} -
{1\over g} \partial_l P_{kl}^\ast\,.
\ee
Up to divergence terms this vector coincides with the spin density part of 
the Noetherian angular momentum $ S_i (x) := \epsilon_{ijk}A_j^aE_{ak}$
after transformation to the new
variables and projection onto the constraint shell.

Using the representations (\ref{eq:gpottr}) and (\ref{eq:elpotn})
one can easily convince oneself that the
variables \( {Q^\ast}\) and \({P^\ast} \) make no contribution to the
Gauss law constraints (\ref{eq:secconstr})
\be
\Phi_a : = O_{as}[\bar{Q}] \Omega^{-1}_{\ sj}
\overline{P}_j = 0~.
\label{eq:4.54}
\ee
The equivalent set of constraints 
\be
\overline{P}_a   = 0
\ee
is Abelian due to the canonical structure of the new variables.
After having rewritten the model in terms of the new canonical coordinates
and after the Abelianization of the Gauss law,
the construction of the unconstrained Hamiltonian system
is straightforward. In all expressions we can simply put
\(\overline{P} = 0\).
In particular, the Hamiltonian in terms of the unconstrained
canonical variables \(Q^\ast\) and \(P^\ast\)
can be represented by the sum of three terms
\be \label{eq:uncYME}
H  =
\frac{1}{2} \int d^3{x} 
\biggl[\,
\mbox{Tr}(P^\ast)^2 + \mbox{Tr}(B^2(Q^\ast))
\ + {1\over 2}{\vec E}^2(Q^\ast,P^\ast)
\biggr].
\ee
The first term is the conventional quadratic ``kinetic'' part,
the second the trace of the square of the non-Abelian magnetic field
\be
B_{sk}(Q^{\ast})= \epsilon_{klm}
(\partial_l Q^{\ast}_{sm} +
{g\over 2}\epsilon_{sbc} \, Q^{\ast}_{bl}Q^{\ast}_{cm})~.
\ee
The third term in the Hamiltonian is the square of the antisymmetric
part ${\vec E}$ of the electric field (\ref{eq:elpotn}) after projection 
onto the constraint surface and is given as the solution of the
partial differential equations
\be
\label{vecE}
{}^\ast D_{ls}E_{s} (Q^{\ast})= g{\cal S}_l 
\ee
 It describes a nonlocal interaction of spin densities
(\ref{eq:spin}). 

The electric field ${\vec E}$ 
can be expanded $\label{vecE1} E_{s} = \sum_{n=0}^{\infty} E^{(n)}_{s}$
in $1/g$, with the zeroth order term 
\be  
E^{(0)}_{s}=\gamma^{-1}_{sk}\epsilon_{klm}
\left(P^\ast Q^{\ast}\right)_{lm}~,
\ee 
where $\gamma_{ik} := Q^\ast_{ik}-\delta_{ik} \mbox{Tr}(Q^\ast)$.
The first order term is determined from the corresponding zeroth order 
term as
\be
E^{(1)}_{s} := {1\over g}
\gamma^{-1}_{sl}\left[(\mbox{rot}\ {\vec {E}}^{(0)})_l 
-\partial_k P^\ast_{kl}\right]~.
\ee
The higher terms are obtained via the simple recurrence relations
\footnote{These expressions can be rewritten 
in terms of the covariant curl operation 
$
\mbox{curl}\,S(e_i, e_j) := \big< \nabla_{e_i} S,  e_j \big>  - 
\big< \nabla_{e_j} S,  e_i \big>  
$
using the basis $e_i : = \left (\gamma^{1/2}\right)_{ij} 
{\partial}_j$  such that $\gamma_{ij}: = \big< e_i,  e_j \big>$. 
}
\be
\label{vecE2} 
E^{(n+1)}_{s} := {1\over g}
\gamma^{-1}_{sl}(\mbox{rot}\ {\vec {E}}^{\ (n)})_l
\ee
The initial gauge fields $A_i$ transform as vectors under spatial rotations.
From the Noetherian expression of the total angular momentum in terms of
the physical fields  (neglecting surface terms)
\be
\label{IQP}
I_i = \int d^3 x~\epsilon_{ijk}
\left((Q^\ast P^\ast)_{jk}
+ {1\over g} x_k \mbox{Tr}\left( P^\ast\partial_j Q^\ast \right)\right)~,
\ee
we find that the matrix fields $Q^\ast$ and $P^\ast$ transform as 
second rank tensors under spatial rotations.
Any such tensor can be decomposed into its
irreducible components, one spin-0 and the five components of a spin-2 field
by extraction of its trace \cite{Brink}.
Decomposing the symmetric matrix $Q^\ast $ into  
the irreducible representations of the S0(3) group
\be
\label{QY}
Q^\ast_{ij}(x) = {1\over \sqrt{2}}{Y}_A(x)\  T^{A}_{ij} 
+{1\over \sqrt{3}} \Phi(x)\ I_{ij}
\ee  
with the field \(\Phi \) proportional to the trace of $Q^\ast$ 
as spin-0 field and the five-dimensional spin-2 vector $Y(x)$ with
components $Y_A$ labeled by the value of spin   
projection on the $z$- axis $ A= \pm 2, \pm 1, 0 $.
\footnote{
For the lowering and raising of the indices of 
5-dimensional vectors the  metric tensor 
$\eta_{AB}= (-1)^A\delta_{A,-B}$ is used.}  
$I$ is the $3\times 3$ unit matrix and the five traceless $3\times 3$ 
spin-2 basis matrices ${T}^A$ satisfying the commutator relations
 $[J^0,T^A]_-= A~T^A$ with
the $SO(3)$ generators $(J^a)_{ik}:=i\epsilon_{iak}$ \cite{Brink}.
The canonical conjugate momenta $P_A(x)$ and $P_\Phi(x)$ to the 
fields $Y_A(x)$ and $\Phi(x)$, respectively, are the components
of the corresponding expansion for the $P^\ast$ variable 
\be
\label{PY}
P^\ast_{ij}(x) = {1\over \sqrt{2}} P_A(x)\ T^{A}_{ij} 
+ {1\over \sqrt{3}} P_\Phi(x) \ I_{ij}~.
\ee 
For the magnetic field $B$ we obtain the expansion  
\be
B_{ij}(x) ={1\over \sqrt{2}} {H}_A(x)\ T^{A}_{ij} +
{1\over \sqrt{2}} h_\alpha(x)\ J^{\alpha}_{ij} +{1\over \sqrt{3}} b(x)\ 
I_{ij}
\ee
with the components 
\bea \label{eq:magspin}
&& {H}_A := {1\over 2} c^{(2)}_{A \beta B} \partial_\beta{Y}^B 
+ {g\over \sqrt{3}}\left(\frac{1}{\sqrt{2}}{}^\ast Y_A - 
\Phi {Y}_A\right) ~,\\
&& {h}_{\alpha}: =  {1\over 2} d^{(1)}_{\alpha B\gamma} 
\partial_\gamma{Y}^B  + 
\sqrt{2\over 3} \partial_\alpha \Phi~,\\
&& b : =  \frac{g}{\sqrt{3}}({1\over 2} Y_A Y^A - {\Phi}^2)~.
\eea
The structure constants $c^{(2)}_{A\beta C}$ 
and $d^{(1)}_{\alpha B \gamma }$ are defined via the algebra 
$[T_{A},T_{B}]_{-} = {c}^{(2)}_{AB\gamma}J^{\gamma}$
and
$[{J}_{\alpha},T_{B}]_{+} =
d^{(1)}_{\alpha \gamma B} {J}^\gamma$ 
respectively,
and the five-dimensional vector 
\be
{}^{\ast}Y_{C}:=  d^{(2)}_{{CAB}} {Y}^A{Y}^B
\ee
via the structure constants $d^{(2)}_{ABC}$ from
$[T_{A},T_{B}]_{+} =
\frac{4}{3}\eta_{AB} I
+ {2\over \sqrt{3}}{d}^{(2)}_{ABC}T^C $.

Note that for a complete investigation of the 
transformation properties
of the reduced matrix field $Q^\ast$ under the whole Poincar\'e group
it is necessary also to include the Lorentz transformations.
But we shall limit ourselves here to the isolation of the scalars under
spatial rotations and can 
treat $Q^\ast$ in terms of ``nonrelativistic  spin-0 and spin-2
fields'', in accordance with the conclusions obtained in the work 
\cite{Faddeev79}.

In summary, we have shown  how to 
project SU(2) Yang-Mills theory onto the constraint 
shell defined by the Gauss law.
However, several questions in connection with the global aspects 
of the reduction  procedure are arising at ths point. 
It is well known that the exponentiation of infinitisimal 
transformations generated  by the  Gauss law operator
can lead only to homotopically trivial gauge transformations, 
continuously deformable to unity.
However, the initial classical action is 
invariant under all gauge transformations 
including the homotopically nontrivial ones.
How does this fact reflect itself on the properties of the 
obtained unconstrained theory? 
In order to discuss the global aspects of the Hamiltonian reduction,
we compare the wellknown exact zero energy solution \cite{Loos} of the
Schr\"odinger equation in the extended quantization scheme, 
where the Gauss law is implemented on the quantum level,
with the corresponding solution of the unconstrained Schr\"odinger equation.
For the original constrained system of $SU(2)$ gluodynamics in terms 
of the gauge fields $A_i^a(x)$ this exact but nonnormalizable solution 
$\Psi[A]$, which satisfies both the functional Schr\"odinger equation 
with zero energy eigenvalue and the Gauss law constraints is    
\be
\label{PsiA} 
\Psi[A] = \exp{\left(\pm 8\pi^2 W[A]\right)}~,
\ee
with so-called ``winding number functional \cite{Jackiw} 
$W[A]: = \int d^3x\  K_0(x)$ defined via the zero component of 
the Chern-Simons secondary 
characteristic  class vector \(
K^\mu(A): = -(16\pi^2)^{-1} \epsilon^{\mu\nu\sigma\kappa} 
\mbox{Tr}\left(
F_{\nu \sigma} A_\kappa -\frac{2}{3}gA_\nu A_\sigma A_\kappa
\right)~.
\) 
The winding number functional is known to be invariant under small but
not under large gauge transformations.

In terms of the new variables $Q^\ast$ and 
$\overline{Q}$ the zero component
of the the Chern-Simons vector $K^\mu$ can be written
\bea \label{K_0}
K^0 (A(Q^\ast, \overline{Q}))  &=&
K^0(Q^\ast)  - {g\over 36\pi^2}\epsilon^{ijk}
 \mbox{Tr}\left(\Omega_i\Omega_j\Omega_k\right) \nn\\
& - &{g\over 24\pi^2} 
\epsilon^{ijk} \partial_i\mbox{Tr}\left( Q^j\Omega_k\right) ~. 
\eea
Here we have used the $SU(2)$ matrix $Q^\ast_l:= Q^\ast_{li}\tau_i$ and the 
$SU(2)$ one-form  components \(
g\Omega_i(\overline Q) := U^{-1}(\overline Q)\partial_iU(\overline Q)= 
\Omega_{ls}\tau^s \left(\partial Q_l/ \partial x_i\right)
\)
with the standart relations to the  orthogonal matricies 
$O({\overline Q})$ via $O_{ab}({\overline Q})=
{1\over 2}\mbox{Tr}(U({\overline Q})\tau_a U^T({\overline Q})\tau_b)$.

The wave functional $\Psi[Q^\ast]$ obtained 
from (\ref{PsiA}) by replacing $A$ by $Q^\ast$ is a zero energy eigenstate
of the corresponding unconstrained Hamiltonian (\ref{eq:uncYME}). 
This follows from two 
important properties of the potential terms of the Hamiltonian 
(\ref{eq:uncYME}).
Firstly, the reduced magnetic field $B_{ij}(Q^\ast)$ can be written
as the functional derivative of $W[Q^\ast]$
Furthermore, the nonlocal part of the physical 
electric field  in the  unconstrained Hamiltonian 
annihilates $W[Q^\ast]$  
\be
{\vec  E}^2[Q, \frac{\delta}{\delta Q_{ij}^\ast(x)} ] W[Q^\ast] = 0~.
\ee
Taking into account that 
the magnetic field $B_i = {}^\ast F_{0i}$ satisfies the Bianchi identity
$D_{i}{}^\ast F_{0i} =0$. 

The second and third terms in (\ref{K_0}) are both surface terms.
The third term gives no contribution if we assume the physical variable 
$Q^\ast$ to vanish at spatial infinity.
About the behaviour of the unphysical variables ${\overline Q}_i$ at spatial
infinity we have no information. The 
requirement of the finiteness of the action usually used to fix the
behaviour of the physical fields does not apply for the unphysical 
field ${\overline Q}$. 
Using the usual boundary condition 
$U(\overline Q)\longrightarrow \pm I\ $ at spatial infinity,
the integral over the second term reduces to an integer n
representing the  corresponding winding of the mapping of
compactified three space into $SU(2)$. 

Hence we obtain the relation
\be
\Psi[A]=\exp[\pm {8\pi^2\over g^2}n]\Psi[Q^\ast]
\ee
between the groundstate wave functionals $\Psi[A]$ of the extended 
quantization scheme and the reduced $\Psi[Q^\ast]$.
We find that the winding number of the original gauge field $A$ only
appears as an unphysical normalization prefactor originating 
from the second term in (\ref{K_0}) which depends only on the unphysical 
${\overline Q}_i$. Furthermore we note that the power $8\pi^2 n/ g^2$, 
is the classical Euclidean action of $SU(2)$ Yang-Mills theory of
self-dual fields \cite{BPST} with winding number $n$ .

We are grateful for discussions with  S.A. Gogilidze, D.M. Mladenov,
V.N. Pervushin, G. R\"opke, A.N. Tavkhelidze and J. Wambach.
One of us (A.K.)  acknowledges the Deutsche Forschungsgemeinschaft for 
providing a visting stipend.
His work was also supported  by the Russian Foundation for
Basic Research under grant No. 98-01-00101.
H.-P. P. acknowledges support by the Deutsche Forschungsgemeinschaft
under grant No. Ro 905/11-2 and by the Heisenberg-Landau program .


\end{document}